\documentclass[hyper]{prop2015}
\usepackage[english]{babel}
\input xy
\xyoption{all}
\usepackage{bbm}

\category{Proceedings}
\subtitle{\href{http://www.maths.dur.ac.uk/lms/109/index.html}{LMS/EPSRC Durham Symposium on Higher Structures in M-Theory}}
\title{Categorical Equivalence and the Renormalization Group}
\begin{acknowledgements}
The results summarized in this note reflect papers written in collaboration
with many individuals, and conversations with many more.  As an incomplete
list, we would like to thank especially A.~Caldararu, R.~Donagi, S.~Hellerman,
S.~Katz, and T.~Pantev, with whom most of these results were worked out,
as well as D.~Benzvi, P.~Pandit, and especially T.~Pantev.
for many useful conversations about derived schemes specifically.  
E.S. has been partially supported over
the course of the work reviewed here by a number of NSF grants,
most recently NSF grant PHY-1720321.
\end{acknowledgements}
\author[E. Sharpe]{Eric Sharpe\inst{a,}\footnote{Corresponding author e-mail:~\href{mailto:ersharpe@vt.edu}{\textsf{ersharpe@vt.edu}}}}
\address[1]{Department of Physics, MC 0435, 850 West Campus Drive, Virginia Tech, Blacksburg, VA 24061, USA}

\begin{abstract}
In this article we review how categorical equivalences are realized by
renormalization group flow in physical realizations of stacks, derived
categories, and derived schemes.  We begin by reviewing the physical
realization of sigma models on stacks, as (universality classes of)
gauged sigma models, and look in particular at properties of sigma models
on gerbes (equivalently, sigma models with restrictions on nonperturbative
sectors), and `decomposition,' in which two-dimensional sigma models
on gerbes decompose into disjoint unions of ordinary theories.  We also discuss
stack structures on examples of moduli spaces of SCFTs, focusing on
elliptic curves, and implications of subtleties there for string
dualities in other dimensions.  In the second part of this article,
we review the physical realization of derived categories in terms of
renormalization group flow (time evolution) of combinations of D-branes, 
antibranes, and tachyons.  In the third part of this article, we review how
Landau--Ginzburg models provide a physical realization of derived
schemes, and also outline an example of a derived structure on a
moduli spaces of SCFTs.  
\end{abstract}
\shortabstract
\begin{document}
\maketitle

\section{Introduction}

Over the last twenty years, we have gained a much better appreciation of
how many abstract mathematical concepts play a role in various aspects of
modern physics.
In particular, there seems to be a general story that 
in physical
realizations of categorical structures, notion of homotopy are often
realized by the renormalization group.  We illustrate the relationship
in Table~\ref{table:cat-equiv}.
\begin{table*}
\centering
\begin{proptabular}{cc}{Some illustrations of how categorical equivalences are
realized in physics.}  \label{table:cat-equiv} 
{\bf Math} & {\bf Physics} \\ \hline
{\bf Stacks} (Deligne--Mumford): & \\
presentation of a stack & gauged sigma model \cite{Pantev:2005zs,Pantev:2005rh,Pantev:2005wj} \\
equivalences of presentations & RG \\
& \\
{\bf Derived categories} (of coherent sheaves): & \\
complex of sheaves & branes/antibranes/tachyons \cite{Sharpe:1999qz,Sharpe:2003dr} \\
quasi-isomorphism & RG \\
& \\
{\bf Derived schemes} ( (-1)-shifted symplectic): & \\
presentation of a derived scheme & Landau--Ginzburg model \cite{Joyce:1304.4508,Brav:1305.6302,Ben-Bassat:1312.0090,Pantev:1111.3209} \\
equivalences & RG \\
\end{proptabular}
\end{table*}
We will review and explore this yoga of physical realizations of homotopy
in the rest of this paper.

We begin in Section~\ref{sect:stacks} by describing sigma models on
Deligne--Mumford stacks \cite{Pantev:2005zs,Pantev:2005rh,Pantev:2005wj}.  
We describe how renormalization group flow
realizes equivalences between presentations, and then discuss novel physical
properties of sigma models on special stacks (gerbes),
most importantly, the `decomposition' conjecture relating sigma models on
gerbes to sigma models on disjoint unions of spaces.  
We also discuss four-dimensional sigma models, and the concrete example of
moduli spaces of SCFTs for elliptic curves, explicitly identifying the
Bagger--Witten line bundle, and some implications for string dualities.

In Section~\ref{sect:der-cat} we describe the physical realization of
derived categories \cite{Sharpe:1999qz,Sharpe:2003dr}, 
in terms of systems of branes, antibranes, and
tachyons, and identify localization on quasi-isomorphisms with
renormalization group flow.  As the physical realization of derived categories
has been described in many places, we confine ourselves to a brief
overview.

In Section~\ref{sect:der-scheme}, we outline some physical realizations
of derived schemes implicit
in the mathematics
literature \cite{Joyce:1304.4508,Brav:1305.6302,Ben-Bassat:1312.0090,Pantev:1111.3209}.
Specifically, we discuss how some properties of two-dimensional
Landau--Ginzburg models are encapsulated by derived schemes,
as derived critical loci and derived zero loci, and how
renormalization group flow again realizes equivalences.  We also discuss
derived structures on moduli spaces of SCFTs, and in particular,
walk through how massless spectrum computations in Landau--Ginzburg
orbifolds realize cotangent complex structures.

\section{Sigma models on stacks and gerbes}
\label{sect:stacks}

This section will give
a survey of
work over the last approximately fifteen years, concerning how one
defines the quantum field theory of string propagation on stacks,
and in particular, string propagation on special stacks known as gerbes.

Now, one popular application of gerbes is as a mechanism to globally define
two-form potentials, just as bundles can be thought of as mechanisms to
define gauge fields on compact spaces.  This application is not what
this paper concerns.

Both stacks and, for the purposes of this paper, gerbes can be thought of
as generalized spaces -- pseudo-geo\-metric constructions that are locally,
though not necessarily globally, spaces.  This section outlines work done
to understand to what extent strings can propagate on these
generalized geometries, the definition and properties of quantum field
theories describing strings on a stack or a gerbe. One of the original
motivations was to understand whether compactifications on stacks or
gerbes describe 
new superconformal field theories (SCFTs), new string compactifications.

\subsection{Brief introduction to (Deligne--Mumford) stacks}

There are several ways to define analogues of geometries that do not
involve point/set topology.  For one example, in noncommutative
geometry, one defines a space via the ring of functions on that space.

Briefly, we define a stack via all the maps from other spaces to the stack.

This is nicely set up for sigma models, in which the path integral
sums over maps into the target.  For a stack, in principle the stack
defines the maps into itself.  Examples of stacks include ordinary
spaces, orbifolds, and gerbes.

Let us briefly consider an example to demonstrate a few key points.
Consider the global quotient stack $[X/G]$, where $G$ is finite.
A map $Y \rightarrow [X/G]$ is a pair
\begin{eqnarray}
\lefteqn{
\bigl(  \mbox{principal } G-\mbox{bundle } E \rightarrow Y,
} \nonumber \\ & \hspace*{0.5in} & 
G-\mbox{equivariant map } {\rm Tot}(E) \rightarrow X \bigr).
\end{eqnarray}
In physics, a sigma model on $[X/G]$ coincides with a global orbifold
by $G$:  the bundle $E \rightarrow Y$ defines the twisted section
on worldsheet $Y$, and the map ${\rm Tot}(E) \rightarrow X$ defines
a map from that twisted sector to the covering space.

See for example 
\cite{fantechi2001stacks,Gomez:9911199,vistoli1989intersection,Metzler:0306176,heinloth2005notes,laumon2018champs,Behrend:0605694,Noohi:0503247,Noohi:0710.2615,Noohi:0808.3799} for
introductions to stacks.

\subsection{Sigma models (in two dimensions)}

Now, how can we define the (two-dimensional)
quantum field theory of a nonlinear
sigma model with target an arbitrary (Deligne--Mumford) stack?

Stacks can be locally presented as spaces, so it may be tempting
to imagine `glueing' nonlinear sigma models on various open patches.
Unfortunately, it is not known how to perform such a glueing for a
full quantum field theory.  (Something like this was proposed in
\cite{Witten:2005px} for the perturbative part of a quantum field
theory, but glueing nonperturbative physics is currently unknown.)

Another tempting option is to utilize the fact that stacks can be
presented as groupoids.  It is tempting to then just try to implement
the groupoid relations in physics.  Unfortunately, it is not clear
how to proceed in this fashion either.  To implement the relations
defined by a group, not a groupoid, one gauges the group action,
which requires various ghost and gauge-fixing techniques
ala Faddeev--Popov and Batalin--Vilkovisky.  To implement groupoid
relations would seem to require a significant generalization of
gauging, Faddeev--Popov, and Batalin--Vilkovisky techniques, a generalization
which does not seem to be known, at least to this author, at this time.

Given the constraints above, the reader might well wonder at this point
why one should believe that a quantum field theory for a sigma model
on a stack should exist.  One answer is that since before this work began,
the Gromov--Witten community had been working on (and now possesses) a notion
of Gromov--Witten invariants of stacks \cite{Chen:2000cy,Abramovich:0603151}.  
Now, a notion of Gromov--Witten
theory is neither necessary nor sufficient for the existence of a full
quantum field theory, but it is usefully suggestive.

With all that in mind, the following proposal was made in
\cite{Pantev:2005zs,Pantev:2005rh,Pantev:2005wj}.
Under mild conditions (see e.g. \cite{Totaro:0207210,edidin2001brauer,Kresch:0301249}),
smooth (Deligne--Mumford) stacks
can be presented as a (stacky) global quotient $[X/G]$, for $X$ a space
and $G$ a group.  $G$ need not be finite, and need not act effectively.
To such a presentation, we associate a $G$-gauged sigma model on $X$.
(Since $G$ need not be finite, this class includes both orbifolds as well
as more general gauge theories; since $G$ does not necessarily act effectively,
it also probes classes of gauge theories not considered prior to 
\cite{Pantev:2005zs,Pantev:2005rh,Pantev:2005wj}.)

Different presentations of the same stack can yield very different 
quantum field theories.  As a simple example, consider the following two
supersymmetric theories:
\smallskip
\begin{enumerate}[i)]
\item the orbifold $[ {\mathbbm C}^2 / {\mathbbm Z}_2 ]$, a ${\mathbbm Z}_2$
gauge theory with two free chiral superfields,
\item a $U(1)$-gauged supersymmetric sigma model with target 
\begin{equation}
X \: = \: \frac{ {\mathbbm C}^2 \times {\mathbbm C}^{\times} }{ {\mathbbm Z}_2 },
\end{equation}
where the generator of ${\mathbbm Z}_2$ acts as
\begin{equation}
(x,y,t) \in {\mathbbm C}^2 \times {\mathbbm C}^{\times} \: \mapsto \:
(-x,-y,-t),
\end{equation}
and the $U(1)$ acts only on the ${\mathbbm C}^{\times}$ factor.
\end{enumerate}
\smallskip
The first of these examples defines a conformal field theory;
the second, because of D-terms in the gauge action, is not conformal.
These are therefore two different quantum field theories, but we expect
that the second theory flows in the IR to the first.

Thus, we cannot associate gauged sigma models themselves to stacks,
but must do something a shade more subtle.
To be precise, we associate
stacks to universality classes of renormalization group flow of such gauged
sigma models.  Put another way, physical realizations of
different (physically-realizable) presentations of stacks are 
related by renormalization group flow, realizing the first row
of Table~\ref{table:cat-equiv}.

In the remainder of this section, we will describe some interesting examples
and applications of these ideas.

\subsection{Sigma models on gerbes}

Returning to stacks, let us consider the important special case of
gerbes.  Mathematically, a global quotient stack $[X/G]$ will
be a gerbe when a nontrivial subgroup of $G$ acts trivially on $X$
(i.e. the group action is technically ineffective).
A sigma model on a gerbe can be described in several equivalent
ways in two-dimensional quantum field theories:
\smallskip
\begin{enumerate}[i)]
\item a gauged sigma model in which a subgroup of the gauge group $G$
acts trivially on the target $X$,
\item a gauged sigma model with a restriction on nonperturbative sectors,
\item a gauged sigma model 'coupled to a topological field theory.'
\end{enumerate}
\smallskip

For the moment, we will focus on the first description, as a gauge theory
in which a subgroup of the gauge group acts trivially, and later will 
return to the second description in terms of restricted nonperturbative
sectors.

In thinking about that first description, we quickly run into a puzzle:
stacks may remember trivial group actions, but why should a quantum field
theory?  Why in physics is a gauge theory with a trivially-acting
subgroup of the gauge group, any different from a gauge theory in which
one only gauges the effectively-acting coset?

For example, in a $U(1)$ gauge theory, if one decides that all fields have
charges that are multiples of two rather than one, what physical
difference can that make?  It sounds solely like a choice of convention.

In fact, there can be a physical difference, arising solely in how the
nonperturbative sector is defined 
\cite{Pantev:2005zs,Pantev:2005rh,Pantev:2005wj}.

In two-dimensional theories, there are essentially\linebreak three different approaches
to see this distinction.

First, on a compact worldsheet, to specify the matter fields uniquely,
one must specify the vector bundle to which the matter fields couple.
In essence, if the gauge field is associated to a line bundle $L$,
then the unambiguous way to say that a matter field has charge $Q$ is to
say that it is a section of the bundle $L^{\otimes Q}$.
Comparing theories with fields of charge one versus fields of charge two,
the two theories have fields coupling to different bundles, hence
have different zero modes, different anomalies, and different physics.

Second, on a noncompact worldsheet, we can distinguish these cases
using the periodicity of the theta angle.  In two dimensions, the
theta angle acts as an electric field, and its periodicity is determined
by the matter content of the theory.  To be precise, if we build a 
capacitor, then as the theta angle is increased, the field density increases,
and eventually the capacitor will pair-produce matter fields once the
field density is high enough.  One can pair produce arbitrarily massive
fields, even fields with masses above the cutoff scale.  
We can distinguish the `gerbe' theory by adding massive minimally-charged
fields, with charges above the cutoff.  Since their charges are $\pm 1$,
this certainly distinguishes massless states of charge $k>1$ from
massless states of charge $1$, and the theta angle periodicity will
detect their presence, even though their mass is above the cutoff scale.

Third, in either case, one can add defects.  Here, for example,
one can add Wilson lines for fields of charge $\pm 1$ to distinguish
the case of massless fields of charge $k>1$ from massless fields of charge $1$.

Let us consider a concrete example, namely an analogue of the two-dimensional
supersymmetric ${\mathbbm C}{\mathbbm P}^{N-1}$ model.  This is a supersymmetric
gauge theory with one gauged $U(1)$ and $N$ chiral superfields of charge
$1$.  (The chiral superfields behave like homogeneous coordinates on the
projective space.)
The gerby analogue of this theory 
\cite{Pantev:2005zs,Pantev:2005rh,Pantev:2005wj} is a $U(1)$ supersymmetric
gauge theory with $N$ chiral superfields of charge $k$, distinguished from
the charge $1$ case as above.  These nonminimal charges have the following
consequences, among others:
\begin{center}
\begin{proptabular}{cc}{Comparison of ordinary versus gerby supersymmetric
${\mathbbm C}{\mathbbm P}^{N-1}$ models.}
Ordinary ${\mathbbm C}{\mathbbm P}^{N-1}$ & Gerby ${\mathbbm C}{\mathbbm P}^{N-1}$ \\ \hline
Anomalous global $U(1)$s: & \\
$U(1)_A \mapsto {\mathbbm Z}_{2N}$ & $U(1)_A \mapsto {\mathbbm Z}_{2kN}$ \\
A model correlation functions: & \\
$\langle x^{N(d+1)}\rangle = q^d$ & $\langle x^{N(kd+1)} \rangle = q^d$ \\
Quantum cohomology rings: & \\
${\mathbbm C}[x] / (x^N - q)$ & ${\mathbbm C}[x]/(x^{kN} - q)$
\end{proptabular}
\end{center}
Concretely, these two models have different physics.

In passing, the A model correlation functions of the gerby theory
correlate with a different (equivalent) description of these gerby theories:
as theories with restrictions on nonperturbative sectors, one of the
alternative descriptions we mentioned at the start.

Now, restricting nonperturbative sectors violates cluster decomposition,
one of the fundamental axioms of quantum field theory.  Similarly,
if one computes chiral rings and spectra in such theories, one finds
multiple dimension-zero operators, again signaling a violation of
cluster decomposition.

\subsection{Decomposition}

The resolution of this cluster decomposition issue lies in
`decomposition' \cite{Hellerman:2006zs}.  Briefly,
\begin{center}
\begin{tabular}{ccc}
strings on gerbes & = & strings on disjoint unions\\
& &  of spaces.
\end{tabular}
\end{center}
Strings on disjoint unions of spaces also violate cluster decomposition,
but in a manner that is straightforward to understand and control.

This decomposition can take different forms under different circumstances,
but some form of it is commonplace.  We give two families of examples
below:
\smallskip
\begin{enumerate}[i)]
\item Version for nonlinear sigma models on spaces and orbifolds
\cite{Hellerman:2006zs}.  Consider
the global quotient $[X/G]$ where $G$ is an extension
\begin{equation}
1 \: \longrightarrow \: K \: \longrightarrow \: G \: \longrightarrow \: H \:
\longrightarrow \: 1,
\end{equation}
and $K$ acts trivially on $X$.  For simplicity, also assume that the
gerbe is `banded.'  Then, in these circumstances, for $Y = [X/H]$
(the effectively-acting quotient, decomposition predicts
\begin{equation}
{\rm QFT}\left( [X/G] \right) \: = \:
{\rm QFT}\left( \coprod_{ \hat{G} } (Y,B) \right),
\end{equation}
where $\hat{G}$ is the set of irreducible representations of $G$,
and the $B$ field $B$ is determined by the image of the
characteristic class of the gerbe under the map
\begin{equation}
H^2(Y,Z(G)) \: \stackrel{ Z(G) \rightarrow U(1) }{\longrightarrow} \:
H^2(Y,U(1)).
\end{equation}
\item Version for nonabelian two-dimensional gauge theories
\cite{Sharpe:2014tca}.  Briefly, if the matter is invariant under a subgroup
of the center of the gauge group, then the QFT will decompose with
factors differing by discrete theta angles.  For example,
schematically,
\begin{quotation}
pure $SU(2)$ $\: = \:$ pure $SO(3)_+$ $\: + \: $ pure $SO(3)_-$.
\end{quotation}
\end{enumerate}
\smallskip
It should be noted that decomposition will be altered if the original
theory has a nontrivial $B$ field background or nonzero theta angle.
It should also be noted that decomposition is only claimed for
two-dimensional theories, not theories in higher dimensions.

Decomposition has a straightforward understanding in terms of path integrals.
Consider for example a nonlinear sigma model on a gerbe over a space
or orbifold $Y$.  Briefly, summing over nonlinear sigma models on $Y$ with
different $B$ fields projects out many nonperturbative sectors, realizing
the gerbe theory (as a theory with a restriction on nonperturbative
sectors).  Schematically, the path integral of the theory with a
restriction on nonperturbative sectors takes the form \cite{Sharpe:2014tca}
\begin{eqnarray}
\lefteqn{
\int [D \phi] \exp(-S) \sum_B \exp\left( \int \phi^* B \right)
} \nonumber \\
& \hspace*{0.5in} = &
\sum_B \int [D \phi] \exp\left( -S + \int \phi^* B \right),
\end{eqnarray}
which is the path integral for a nonlinear sigma model on a disjoint
union of spaces with variable $B$ fields.

Let us now consider a more concrete example.  We will recover decomposition
for the case of a global orbifold, describing a ${\mathbbm Z}_2$ gerbe
over $[X / {\mathbbm Z}_2 \times {\mathbbm Z}_2 ]$.  Specifically,
consider $[X/D_4]$, where 
\begin{equation}
1 \: \longrightarrow \: {\mathbbm Z}_2 \: \longrightarrow \: D_4 \:
\longrightarrow \: {\mathbbm Z}_2 \times {\mathbbm Z}_2 \: \longrightarrow \:
1,
\end{equation}
where ${\mathbbm Z}_2$ (equal to the center of $D_4$) acts trivially on
$X$.

In this example, decomposition predicts
\begin{equation}
{\rm CFT}\left( [X/D_4] \right) \: = \:
{\rm CFT}\left( [ X / {\mathbbm Z}_2 \times {\mathbbm Z}_2 ]
\coprod
 [ X / {\mathbbm Z}_2 \times {\mathbbm Z}_2 ]
\right),
\end{equation}
where one of the ${\mathbbm Z}_2 \times {\mathbbm Z}_2$ orbifolds has 
discrete torsion and the other does one.

Next, we shall see how to recover decomposition at the level of string
one-loop partition functions.  Label the group elements as follows:
\begin{eqnarray}
D_4 & = & \{ 1, z, a, b, az, bz, ab, ba = abz \}, \\
{\mathbbm Z}_2 \times {\mathbbm Z}_2 & = &
D_4/{\mathbbm Z}_2 \: = \:
\{ 1, \overline{a}, \overline{b}, \overline{ab} = \overline{ba} \},
\end{eqnarray}
where $z \in D_4$ generates the center (${\mathbbm Z}_2$), and we use
bars to denote cosets, e.g. $\overline{a} = \{a, az \}$.
The string one-loop partition function has the form
\begin{equation}
Z\left( [X/D_4] \right) \: = \:
\frac{1}{|D_4|} \sum_{gh=hg} Z_{g,h},
\end{equation}
where each $Z_{g,h}$ is a sum over maps with boundary conditions determined
by $g$, $h$.
Note that since $Z_{g,h}$ is determined only by boundary conditions,
each $Z_{g,h}$ for the $D_4$ orbifold is the same as a $Z_{\overline{g},
\overline{h}}$ of a ${\mathbbm Z}_2 \times {\mathbbm Z}_2$ orbifold,
appearing with multiplicity $| {\mathbbm Z}_2 |^2 = 4$,
except for the ${\mathbbm Z}_2 \times {\mathbbm Z}_2$ twisted
sectors defined by the pairs $(\overline{a}, \overline{b})$,
$(\overline{a}, \overline{ab})$, and
$( \overline{b}, \overline{ab})$, which do not have a lift to
$D_4$.  (The preimages of the group elements in $D_4$ do not commute.)

As a result of the counting above, we see
\begin{eqnarray}
\lefteqn{
Z\left( [X/D_4] \right)
} \nonumber \\
 & = &
\frac{ | {\mathbbm Z}_2 \times {\mathbbm Z}_2 | }{ | D_4 | }
| {\mathbbm Z}_2 |^2 \cdot 
\nonumber \\
& & \cdot
\left(
Z\left( [X/{\mathbbm Z}_2 \times {\mathbbm Z}_2] \right) \: - \:
\left( \mbox{some twisted sectors} \right) \right),
\\
& = &
2 \left(
Z\left( [X/{\mathbbm Z}_2 \times {\mathbbm Z}_2] \right) \: - \:
\left( \mbox{some twisted sectors} \right) \right).
\end{eqnarray}
Discrete torsion acts as a sign on the omitted twisted sectors above
of the ${\mathbbm Z}_2 \times {\mathbbm Z}_2$ orbifold, hence
\begin{equation}
Z\left( [ X/D_4] \right) \: = \:
Z \left( [ X / {\mathbbm Z}_2 \times {\mathbbm Z}_2 ] \coprod
[ X / {\mathbbm Z}_2 \times {\mathbbm Z}_2 ] \right)
\end{equation}
with discrete torsion in one component, consistent with the prediction of
decomposition.

We list here two sets of applications of decomposition.
\smallskip
\begin{enumerate}[i)]
\item Gromov--Witten theory.  In particular, one prediction of
decomposition is that Gromov--Witten invariants of gerbes match
Gromov--Witten invariants of disjoint unions of spaces.
This has since been proven in work of H.-H.~Tseng, Y.~Jiang, and
collaborators in e.g. \cite{Andreini:0812.4477,Andreini:0905.2258,Andreini:0907.2087,Tseng:0912.3580,Gholampour:1001.0435,Tang:2010pf}.
\item Phases of gauged linear sigma models (GLSMs).\linebreak
Phases of certain GLSMS, which were previously obscure, now have a 
solid understanding utilizing decomposition.  The prototype 
\cite{Caldararu:2007tc} is the
Landau--Ginzburg orbifold point of the GLSM for
a complete intersection of $n$ quadrics
in a projective space ${\mathbbm P}^{2n-1}$.  Briefly, utilizing
decomposition, this Landau--Ginzburg point can be interpreted as a 
(possibly noncommutative resolution of a) branched double cover of
${\mathbbm P}^{n-1}$, branched over a degree $2n$ locus.
The double cover structure is a local application of decomposition for
a ${\mathbbm Z}_2$ gerbe.
For other applications see e.g. 
\cite{Hori:2011pd,Halverson:2013qca,Hori:2016txh,Wong:2017cqs,Chen:2018qww}.
\end{enumerate}

\subsection{Four-dimensional sigma models on stacks}

So far we have focused on two-dimensional nonlinear sigma models,
that is, QFTs on two-dimensional spaces with targets of targets of
possibly other dimension.

In principle, we can also consider four-dimensional low-energy effective
nonlinear sigma models, that is, quantum field theories on four-dimensional
spaces with targets of possibly other dimension.  These arise in e.g. 
four-dimensional supergravity theories, describing the space of scalar field
vevs, and analogous considerations apply there.  (See \cite{Hellerman:2010fv}
for a more detailed discussion.)

A four-dimensional nonlinear sigma model on a gerbe can be distinguished
from an ordinary nonlinear sigma model in much the same way as 
two-dimensional cases:
\smallskip
\begin{enumerate}[i)]
\item Compact four-dimensional spaces:  to specify matter fields, one must
specify bundles, and different bundles give rise to different anomalies and
zero modes, and hence different physics, just as in two dimensions.
\item Noncompact four-dimensional spaces:
In four dimensions, the theta angle no longer acts like an electric
field as it does in two dimensions.  However, instead of using theta angles,
we can use charged Reissner-Nordstrom black holes.  As before, to distinguish
nonminimal charges from minimal charges, we add massive minimally-charged
fields.  These fields can be emitted in Hawking radiation from black holes.
To be specific, consider a four-dimensional $U(1)$ gauge theory with
massless fields whose charges are multiples of $k$.
If there are massive minimally-charged fields, then a charged
black hole can Hawking radiate down to charge $1$, whereas if there are no
massive minimally-charged fields, then it can only Hawking radiate to charge
$k$.
\item Defects:  we can for example add Wilson lines of charges that are
only well-defined in certain theories, thereby distinguishing between
different four-dimensional theories with the same low-energy matter content.
\end{enumerate}
\smallskip

Next, we will consider some concrete examples.

In a perturbative string compactification, the low-energy four-dimensional
supergravity contains a low-energy effective nonlinear sigma model whose
target is the moduli stack of Calabi--Yau's, or more generally, a moduli
space of two-dimensional SCFTs.  
So, concrete examples of four-dimensional sigma models on gerbes can be
found in moduli spaces of Calabi--Yau's.  To be concrete, we will consider
moduli spaces of elliptic curves next.

\subsection{Moduli stacks of elliptic curves, the metaplectic group,
and string dualities}   \label{sect:mod-ell-curves}

Moduli spaces of elliptic curves are often presented as $PSL(2,{\mathbbm Z})$
quotients of the upper half plane.
The action on the upper half plane is given by
\begin{equation}
\tau \: \mapsto \: \frac{ a \tau + b}{c \tau + d},
\end{equation}
for
\begin{equation}
\left[ \begin{array}{cc}
a & b \\ c & d \end{array} \right] \: \in \: SL(2,{\mathbbm Z}).
\end{equation}
The center diag$(-1,-1) \in SL(2,{\mathbbm Z})$ acts trivially
on $\tau$, hence the action descends to a $PSL(2,{\mathbbm Z})$ quotient.

However, the $PSL(2,{\mathbbm Z})$ quotient loses information.
We can see this as follows.  Describe the elliptic curve as
${\mathbbm C} / (\tau {\mathbbm Z} + {\mathbbm Z})$, and let $z$ be an
affine coordinate on ${\mathbbm C}$.  Then, under $SL(2,{\mathbbm Z})$
(see e.g. \cite{Hain:0812.1803}[Section 2.3]),
\begin{equation}
(\tau, z) \: \mapsto \: \left( \frac{ a \tau + b}{c \tau + d},
\frac{z}{c \tau + d} \right),
\end{equation}
hence under $SL(2,{\mathbbm Z})$,
\begin{equation}
{\rm d}z \: \mapsto \: \frac{{\rm d}z}{c \tau + d}.
\end{equation}
In particular, the central element diag$(-1,-1) \in SL(2,{\mathbbm Z})$
acts trivially on $\tau$ but sends ${\rm d}z \mapsto - {\rm d}z$,
so the holomorphic top-form on an elliptic curve is not well-defined
in the $PSL(2,{\mathbbm Z})$ quotient of the upper half plane.

To build a line bundle over the moduli space encoding a family of 
canonical line bundles (technically, the Hodge line bundle), 
we must take the moduli space to be
$[ {\mathfrak h} / SL(2,{\mathbbm Z}) ]$, for ${\mathfrak h}$ the
upper half plane, instead of $[ {\mathfrak h}/PSL(2,{\mathbbm Z}) ]$.

In physics, the holomorphic top-form corresponds to the spectral flow
operator of the $N=2$ algebra, so in order to construct a moduli space
of SCFTs for sigma models on elliptic curves, over which one has a family
of spectral flow operators, one must work with the $SL(2,{\mathbbm Z})$
quotient, which is a ${\mathbbm Z}_2$ gerbe over the $PSL(2,{\mathbbm Z})$
quotient.

Mathematically, it turns out that the moduli space ${\cal M}_{1,1}$
of elliptic curves is identified with the $SL(2,{\mathbbm Z})$ quotient:
\begin{equation}
{\cal M}_{1,1} \: = \: 
[ {\mathfrak h}/SL(2,{\mathbbm Z})].
\end{equation}
In addition, since there is a naturally-defined line bundle of
spectral flow operators (holomorphic top-forms), this is also at least
part of what we need for a good moduli space of SCFTs.

However, it turns out that to get a good moduli space of SCFTs (of sigma
models on elliptic curves), we need more.
The issue is that the chiral Ramond vacuum is not well-defined under
either the $PSL(2,{\mathbbm Z})$ or $SL(2,{\mathbbm Z})$ quotients.

Specifically, under $SL(2,{\mathbbm Z})$,
\begin{equation}   \label{eq:vacuum-transformation}
(\tau, z, | 0 \rangle ) \: \mapsto \:
\left( \frac{a \tau + b}{c \tau + d}, \frac{z}{c \tau + d},
\pm \frac{ | 0 \rangle }{\sqrt{ c \tau + d}} \right).
\end{equation}
In principle, this is a consequence of the fact that the Fock vacuum
in a sigma model with target $X$ couples to the pullback of $K_X^{1/2}$.
(This is implicit in the NSR formalism \cite{Sharpe:2013bwa}, 
and follows ultimately from the
fact that on a K\"ahler manifold $X$, the spinor bundle can be described
as \cite{lawson2016spin}[Equation (D.16)]
\begin{equation}
\wedge^{\bullet} TX \otimes \sqrt{K_X},
\end{equation}
with the wedge interpreted as the complex exterior\linebreak power, not the real
exterior power.
In the worldsheet realization, the $\wedge^{\bullet} TX$ is formed
by multiplying a Ramond vacuum by various worldsheet fermions $\psi$,
and the Fock vacuum itself corresponds to $\sqrt{K_X}$.)

We can understand the role of the central element of $SL(2,{\mathbbm Z})$
more concretely as follows.  First, in a sigma model on $T^2$,
there is a single complex fermion $\psi$, and in a chiral Ramond sector,
strictly speaking there are two chiral Ramond vacua $| \pm \rangle$,
which we define as follows:
\begin{equation}
\psi | + \rangle \: = \: 0, \: \: \:
\psi | - \rangle \: = \: | + \rangle, \: \: \:
\overline{\psi} | - \rangle \: = \: 0.
\end{equation}
Ultimately because these vacua have fractional charges, under
the transformation\footnote{Strictly speaking, there are two possible
${\mathbbm Z}_2$ orbifolds, which can be distinguished formally by
${\mathbbm Z}_2$ and ${\mathbbm Z}_2 (-)^F$.
In writing the above, we are choosing one of these, consistent with the
action on target-space spinors -- consistent with the fact that a 
rotation in space does not return a spinor to itself, as Spin is a
double cover of $SO$.
} $\psi \mapsto - \psi$, these vacua transform as
\begin{equation} 
| \pm \rangle \: \mapsto \: \pm \exp\left( i \pi / 2 \right) | \pm \rangle.
\end{equation}
This is consistent with the transformation~(\ref{eq:vacuum-transformation})
above, for the central element of $SL(2,{\mathbbm Z})$.

In any event, this means there is a sign ambiguity in the action of
$SL(2,{\mathbbm Z})$ on the Fock vacuum; the action of $SL(2,{\mathbbm Z})$
is not well-defined.  To get a well-defined action, we must replace
$SL(2,{\mathbbm Z})$ by a ${\mathbbm Z}_2$ extension.
In fact, it can be shown that \cite{Gu:2016mxp} chiral Ramond vacua couple
to a naturally-defined line bundle over the quotient
$[ {\mathfrak h} / Mp(2,{\mathbbm Z}) ]$,
where $Mp(2,{\mathbbm Z})$ is the metaplectic group, the unique nontrivial
${\mathbbm Z}_2$ extension of $SL(2,{\mathbbm Z})$.

We conjecture \cite{Gu:2016mxp} that the moduli space of (complex structures
on) SCFTs for sigma models on elliptic curves is
$[ {\mathfrak h} / Mp(2,{\mathbbm Z}) ]$.

In fact, we have also argued \cite{Pantev:2016nze} that the metaplectic group
arises elsewhere in string dualities:
\smallskip
\begin{enumerate}[i)]
\item T-duality of $T^2$:  for essentially the same reasons as above,
one should replace the $SL(2,{\mathbbm Z})$ factors in the T-duality group
\begin{equation}
SO(2,2;{\mathbbm Z})
\end{equation}
by $Mp(2,{\mathbbm Z})$.
\item Ten-dimensional IIB S-duality:  because of analogous sign ambiguities
in S-duality actions on ten-dimen\-sional fermions, one should also replace\linebreak
$SL(2,{\mathbbm Z})$ by $Mp(2,{\mathbbm Z})$.  (This has also been observed
by D.~Morrison.)
\item M theory on $T^2$:  one can argue that the action of the mapping class
group on fermions is $Mp(2,{\mathbbm Z})$.
\item U-duality in nine dimensions:  $Mp(2,{\mathbbm Z})$ instead of
$SL(2,{\mathbbm Z})$.
\end{enumerate}
\smallskip
These dualities are interrelated:  the U-duality group in nine dimensions can
be understood either from M theory on $T^2$ or from ten-dimensional IIB
S-duality, and so it is a nontrivial consistency check that these different
duality groups match.

\subsection{Open questions}

A natural question to ask is, which moduli stack is sensed by 
defects in ten-dimensional IIB?  The stack $[ {\mathfrak h} / PSL(2,{\mathbbm Z}) ]$,
$[ {\mathfrak h}/ SL(2,{\mathbbm Z}) ]$, or
$[ {\mathfrak h}/ Mp(2,{\mathbbm Z}) ]$?
In F-theory compactifications, there is a defect which senses the center of
$SL(2,{\mathbbm Z})$, corresponding to a Kodaira fiber $I_0^*$, or a
D7-brane on an O7-plane.  One can then ask whether there is a D7-brane
configuration that senses the ${\mathbbm Z}_2$ specific to $Mp(2,{\mathbbm Z})$.

Another 
natural question for future work is whether there is any three-dimensional
analogue of decomposition.  One does not expect the three-dimensional theory
to decompose as a disjoint union of quantum field theories, but there might
be some sort of decomposition for certain classes of defects within the
theory, for example.  A first pass at answering this question
is implicit in \cite{Aharony:2017adm}.

Finally, let us conclude with a conjecture motivated by 
recent work.  The Bagger--Witten and Hodge line bundles over moduli
stacks of Calabi--Yau's are known in
only a few examples \cite{Gu:2016mxp,Donagi:2017mhd},
but in those examples, they are nontrivial but generate a finite subgroup
of the Picard group.  We conjecture that this is true more generally:
\begin{quotation}
{\it Conjecture:  over any Calabi--Yau moduli space, the Bagger--Witten and Hodge line bundles are holomorphically nontrivial but admit flat connections.}
\end{quotation}
This is a refinement of ideas expressed in \cite{Gomis:2015yaa},
which gave a physical argument that the Bagger--Witten line bundle should be
flat (but did not require nontriviality).
This would also be an analogue of, and related to, the weak gravity conjecture
\cite{ArkaniHamed:2006dz}) for existence of UV completions of four-dimensional
supergravity theories.

\section{Derived categories}  \label{sect:der-cat}

Of the various topics discussed in this overview, the physical realization
of derived categories, as combinations of branes, antibranes, and tachyons,
is relatively well-known, so we will be comparatively brief.
(See e.g. \cite{Sharpe:2003dr} for a more detailed review of the
physical realization, and \cite{Caldararu:0501094,Thomas:1999ic,Weibel:1994aa,hartshorne1966residues} for
more information on the mathematics of derived categories.)

The physical realization of derived categories, first described in
\cite{Sharpe:1999qz}, was originally motivated by two separate developments.
\smallskip
\begin{enumerate}[i)]
\item Kontsevich's homological mirror symmetry \cite{Kontsevich:9411018},
relating the derived category of coherent sheaves of one Calabi--Yau to
a derived Fukaya category of the mirror Calabi--Yau.
When this was originally proposed, mirror symmetry was only understood
as a relationship between closed string theories, and the physical meaning,
if any, of derived categories of coherent sheaves and derived Fukaya
categories, was unknown.  Part of the motivation of \cite{Sharpe:1999qz} was
to find a physical underpinning for homological mirror symmetry.
\item Sen's work on antibranes 
\cite{Sen:1998rg,Sen:1998ii,Sen:1998sm,Sen:1998ex}.  
Sen introduced the
idea of antibranes and pertinent facts about brane-antibrane annihilation,
which were interpreted by Witten mathematically in terms of K-theory
\cite{Witten:9810188}.  However, this work only kept track of smooth information,
and so another motivation was to find a holomorphic analogue,
which in physics could keep track of e.g. information about
connections (morally) on smooth bundles.
\end{enumerate}
\smallskip
Derived categories provided a holomorphic analogue of K-theory,
and an interpretation as some sort of version of tachyon condensation
answered the riddle about the physical meaning of Kontsevich's proposal.

Before talking about derived categories of sheaves, let us first quickly
review the dictionary between ordinary coherent sheaves and D-branes.
Briefly, we know such a dictionary for various special cases.  The most
common case is as follows.  Let $i: S \hookrightarrow X$ be a submanifold
of some Calabi--Yau $X$, with holomorphic vector bundle ${\mathcal E}
\rightarrow S$.  Then, the dictionary \cite{Katz:2002gh} equates
the sheaf $i_* {\mathcal E}$ to a D-brane on $S$ with gauge
bundle ${\cal E} \otimes K_S^{-1/2}$.  The factor of $\sqrt{K_S}$ is ultimately
a reflection of the Freed--Witten anomaly \cite{Freed:1999vc},
as discussed in \cite{Katz:2002gh}, 
and is important in order to match open string
B model boundary chiral rings with Ext groups between sheaves.
Another set of known special cases relates structures sheaves of
nonreduced subschemes to
D-branes with nilpotent Higgs vevs \cite{Donagi:2003hh,Cecotti:2010bp}.
Simple statements are not known for other cases (except via projective
resolutions, as we will discuss next.)  The dictionary is summarized
in Table~\ref{table:sheaf-brane-dict}.

\begin{table}[h]
\centering
\begin{proptabular}{cc}{Sheaf / D-brane dictionary.}
\label{table:sheaf-brane-dict}
Sheaf & D-brane \\ \hline
$i_* {\mathcal E}$ & D-brane on $S$ with bundle 
${\mathcal E} \otimes K_S^{-1/2}$ \cite{Katz:2002gh}
\\
nonreduced scheme & Nilpotent Higgs vev, 
T-brane \cite{Donagi:2003hh,Cecotti:2010bp}
\end{proptabular}
\end{table}

So far we have discussed D-branes, but not antibranes,
and the latter are important for the physical realization of
derived categories.  Schematically, a complex of sheaves is mapped
to a collection of branes and antibranes (corresponding to the
sheaves in the complex, with grading mod two
determining the distinction between
brane and antibrane) together with tachyons, which realize the maps
in the complex.  Now, there are subtleties, including both the fact
that we do not know a simple dictionary between all possible sheaves
and D-branes, only certain sheaves, and that whether or not one has
a tachyon in the brane/antibrane spectrum depends upon the difference
in dimensions between the brane and antibrane.

In broad brushstrokes, we deal with these issues as follows.
We replace any complex of sheaves on a Calabi--Yau
by a projective resolution
consisting of locally-free sheaves.  For locally-free sheaves,
we know the corresponding D-branes (which are defined by the
sheaves themselves), and physically there exist tachyons between
all branes and 
antibranes here since they all have the same dimension.

Let us now illustrate these ideas in greater detail.
Boundary actions for brane, antibrane, tachyon systems were
constructed in \cite{Hori:2000ic}[Section 5.1.2],
\cite{Kraus:2000nj}[Section 4],
\cite{Takayanagi:2000rz}[Section 2],
and \cite{Herbst:2008jq}, and take the form
\begin{eqnarray}
\lefteqn{
\int_{\partial \Sigma} {\rm d}x \Big[ 
\overline{\eta} d \eta + i \psi^i \left(\partial_i P\right) \eta + i
\psi^{\overline{\imath}} \left( \partial_{\overline{\imath}}
\overline{P} \right) \overline{\eta}
}\notag \\
& \hspace*{0.5in} & + i \psi^i \left( \partial_i Q\right) \overline{\eta}
+ i \psi^{\overline{\imath}} \left( \partial_{\overline{\imath}} 
\overline{Q} \right) \eta
\\
& \hspace*{0.5in} & - i |P|^2 - i |Q|^2 \Big],\notag
\end{eqnarray}
where $\psi^i = \psi_+^i + \psi_-^i$ is the restriction of the
bulk worldsheet fermions to the boundary, and $\eta$, $\overline{\eta}$ are
fermions living only on the boundary.  There are two vector bundles
(associated with the branes and antibranes), which we will label
${\cal E}_{0,1}$.  The boundary fermion $\eta$ couples to
${\cal E}_0^* \otimes {\cal E}_1$, and $\overline{\eta}$ couples to
${\cal E}_0 \otimes {\cal E}_1^*$.  The field $P$ is a section of
${\cal E}_0^* \otimes {\cal E}_1$, and $Q$ is a section of
${\cal E}_0 \otimes {\cal E}_1^*$.  Under a supersymmetry transformation,
for which the boundary fermions transform as
\begin{eqnarray}
\delta \eta & = & -i \overline{P} \alpha -i Q \tilde{\alpha}, \\
\delta \overline{\eta} & = & -i P \tilde{\alpha} -i \overline{Q} \alpha,
\end{eqnarray}
the supersymmetry variation of the boundary action takes the form
\begin{equation}
\int_{\partial \Sigma} \Big[
- \alpha \psi^{\overline{\imath}} \partial_{\overline{\imath}} 
\left( \overline{P} \overline{Q} \right) -
\tilde{\alpha} \psi^i \partial_i \left( P Q \right) \Big].
\end{equation}
If the bulk worldsheet theory had a superpotential, as in a Landau--Ginzburg
model, then one could solve the 
Warner problem \cite{Warner:1995ay,Kapustin:2002bi} by requiring
$PQ = W {\rm Id}$, up to a constant shift, which leads to matrix
factorizations.  If the bulk theory is just a 
nonlinear sigma model without superpotential, 
then instead we require $PQ = 0$ (up to a constant).

Now, the bulk worldsheet sigma model also has a pair of $U(1)_R$
symmetries, acting on the left- and -right-moving worldsheet fermions,
which in principle on the boundary restrict to a common $U(1)_R$.
To recover the grading implicit in a derived category, we require that the
bundles ${\cal E}_{0,1}$ be $U(1)_R$ equivariant, and that the maps
$P$, $Q$ each have $U(1)_R$ charge $+1$.  The fermions $\psi^i$ then have
$U(1)_R$ charge $-1$.  (In the case of a matrix factorization,
this is consistent with the convention that the worldsheet superpotential have
$U(1)_R$ charge two.  See e.g. \cite{Segal:2010cz,Addington:2012zv}
for further information on the $U(1)_R$ action for matrix factorizations.)

Now, in a (bulk) nonlinear sigma model without superpotential,
the $U(1)_R$ acts only on the fermions, not the bosons, so 
we are requiring that the bundles ${\cal E}_{0,1}$ be equivariant with
respect to a group that acts trivially on the space over which they are
defined.  In such a case, the action is ${\mathbbm Z}$-graded.  
In general, ${\cal E}_0$ and ${\cal E}_1$ may decompose, so we write
\begin{equation}
{\cal E}_0 \: = \: \oplus_i {\cal A}_i, \: \: \:
{\cal E}_1 \: = \: \oplus_j {\cal B}_j.
\end{equation}
Each summand ${\cal A}_i$, ${\cal B}_j$ can have a different integral
weight under $U(1)_R$, so without loss of generality, we will identify
the $U(1)_R$ charge with the integer index -- meaning, for example,
we will take ${\cal A}_i$ to have $U(1)_R$ weight $i$
and ${\cal B}_j$ to have $U(1)_R$ weight $j$.
Then, since $P$ is a section of ${\cal E}_0^* \otimes {\cal E}_1$,
it defines a set of maps
\begin{equation}
P_i: \: {\cal A}_i \: \longrightarrow \: B_{i+1},
\end{equation}
and similarly $Q$ defines a set of maps
\begin{equation}
Q_i: \: {\cal B}_i \: \longrightarrow \: {\cal A}_{i+1}.
\end{equation}
Since $PQ = 0$, we see that these maps form a complex
\begin{equation}
\cdots \: \longrightarrow \: {\cal A}_i \: \stackrel{ P_i }{\longrightarrow}
\: {\cal B}_{i+1} \: \stackrel{ Q_{i+1} }{\longrightarrow} \:
{\cal A}_{i+2} \: \stackrel{ P_{i+2} }{\longrightarrow} \: 
{\cal B}_{i+3} \: \stackrel{Q_{i+3}}{\longrightarrow} \: \cdots.
\end{equation}
(In principle, we have sufficient data for two complexes of this form.
However, for simplicity, we assume here that, for example,
${\cal A}_{\rm odd}$ and ${\cal B}_{even}$ all vanish, which also allows
us to cleanly distinguish branes from antibranes (one is encoded by the
${\cal A}_i$, the other by the ${\cal B}_i$).
In this fashion, we see that $U(1)_R$-equivariant boundary data
defines a complex of bundles.

Now, there are essentially two classes of isomorphisms between the 
brane-antibrane systems above.  The first is defined by homotopies of
complexes.  Maps between complexes can again be physically realized in terms
of tachyons, and it can be shown (see e.g. \cite{Aspinwall:2001pu}) 
that chain-homotopic
maps are BRST-equivalent in that realization.

The second class of isomorphisms we shall discuss is that of
quasi-isomorphisms.
Quasi-isomorphisms between complexes are realized in physics by
renormalization group flow (realizing an entry in
Table~\ref{table:cat-equiv}).  This is straightforward to outline in an
example.  Consider a brane described by the structure sheaf ${\cal O}$,
and an antibrane on some ideal sheaf ${\cal O}(-D)$, for some divisor
$D$, together with a tachyon map corresponding to the inclusion
${\cal O}(-D) \rightarrow {\cal O}$.  Physically, one expects that such 
a brane-antibrane collection should mostly annihilate, physically evolving
over time into a single brane supported along the divisor $D$.
Such time-evolution in spacetime corresponds to renormalization group flow
on the worldsheet, and mathematically we are identifying
\begin{equation}
0 \: \longrightarrow \: {\cal O}(-D) \: \longrightarrow \: {\cal O}
\: \longrightarrow \: 0
\end{equation}
with
\begin{equation}
0 \: \longrightarrow \: {\cal O}_D \: \longrightarrow \: 0.
\end{equation}
These two complexes are quasi-isomorphic to one another, and so we
see in this example that quasi-isomor\-phism is the mathematical
implementation of renormalization group flow.

So far we have just discussed the physical realization of derived
categories.  Perhaps the most well-known application is to stability
questions.  That is beyond the scope of this review;
we instead refer interested readers to
e.g.
\cite{Douglas:2000gi,Bridgeland:0212237}.

\section{Derived schemes}
\label{sect:der-scheme}

In Section~\ref{sect:der-cat}, 
we saw one realization of complexes -- in terms of
branes, antibranes, and tachyons.  In principle, there are other
places where complexes enter physics.  One example is via Yukawa couplings
in two-dimensional Landau--Ginzburg theories.
Another is in massless spectrum computations in string compactifications.
Typically, we describe massless spectra and BRST cohomology as the
cohomology of some complex.  That complex has meaning inside the
quantum field theory -- each element of the complex represents some set of
states or operators, 
which may or may not be massless or BRST closed, but which can be
explicitly represented within the QFT.
Mathematically, the complexes above can be interpreted in terms of
derived geometry.
In this section we will outline these two examples, their relation
to derived geometry, and their roles
in two- and four-dimensional theories, in close analogy with 
Section~\ref{sect:stacks} on the role of stacks in physics.

Now, to be clear, we do not claim these are the only places where
complexes or derived geometry enter physics, and indeed, one suspects
there are many places in physics where derived geometry can play
a role.
Other physical realizations of derived geometry in different contexts
have also appeared in
e.g. \cite{Beem:2018fng,Costello:2011aa,Elliott:2015rja}.

\subsection{Brief introduction to derived geometry}

In a nutshell, in derived schemes, instead of relating spaces
to ordinary algebras of functions, one associates spaces to dg-algebras
of functions.  As a practical matter, this means that a given derived
space can be presented as a variety of different spaces, of potentially
different dimension, all with different analogues of structure sheaves,
analogous to the manner in which a given stack can have a variety of
different presentations.

One essential aspect of derived geometry is the `cotangent complex,'
often denoted ${\mathbbm L}$,
a complex generalizing the cotangent bundle of smooth manifolds.
Later we will see this arise as the BRST complex in
various situations.
If the derived space is smooth, then the cotangent complex will have
cohomology only in degree 0, and that cohomology sheaf will be precisely
the ordinary cotangent complex.  If the cotangent complex has nonzero
cohomology sheaves in other degrees, then the space is not smooth,
at least in the ordinary sense.
Very readable introductions to the cotangent complex and derived
geometry can be found in \cite{manetticotangent,shincontangent,Malapani:1811.12937}.

Let us briefly describe an example of a prototypical form,
following \cite{manetticotangent}[Example 5.5].
Consider a complete intersection of hypersurfaces in
${\mathbbm C}^n$, defined by the ring
\begin{equation}
A \: = \: {\mathbbm C}[x_1,\cdots,x_n]/(f_1,f_2,\cdots,f_k)
\end{equation}
A dg algebra resolving the ring above is
\begin{equation}
{\mathbbm C}[x_1,\cdots,x_n,y_1,\cdots,y_k]
\end{equation}
with $x$'s of degree $0$, $y$'s of degree $-1$, and differential
\begin{equation}
s(x_i) \: = \: 0, \: \: \:
s(y_j) \: = \: f_j.
\end{equation}
We therefore identify the affine space ${\mathbbm C}^{n+k}$
as an equivalent derived scheme for the
complete intersection, with   
corresponding cotangent complex given by
\begin{equation}
0 \: \longrightarrow \: \oplus_j A {\rm d}y_j \: \stackrel{s}{\longrightarrow}
\: \oplus_i A {\rm d}x_i \: \longrightarrow \: 0,
\end{equation}
where the ${\rm d}x_i$ are in degree $0$ and the ${\rm d}y_j$ in degree $-1$.
The cokernel of the differential, the cohomology at the second step,
is easily checked to be the differential one-forms on the complete
intersection, defined by ${\rm d} x_i$ subject to the equivalence
\begin{equation}  \label{eq:cot-kernel}
 \sum_j \frac{\partial f_j}{\partial x_i}
{\rm d} x_i \: \sim \: 0
\end{equation}
(reflecting the fact that we are restricted to the complete
intersection of hypersurfaces $\{ f_j = 0 \}$).  If the complete
intersection is singular, then at the singularity,
the rank of the cotangent bundle defined
by the constraints above is wrong, and precisely 
in such a case, the differential
$s$ has a nonzero kernel.  Thus, in this example, the cohomology
at degree 0 (identified with the second term) is the cotangent sheaf,
and the complete intersection is singular if and only if there is 
nonzero cohomology at degree $-1$, matching the description of
the cotangent complex above.
(For a compact space (such as a complete intersection in a projective
space), the same story applies patch-by-patch.)

Readers familiar with gauged linear sigma models \cite{Witten:1993yc}
will find the structure above familiar -- a GLSM\footnote{
Strictly speaking, since we are describing affine spaces rather than
projective spaces, nothing is being gauged physically, so the term
GLSM is perhaps not perfectly appropriate.  On the other hand, we can also
consider derived structures for complete intersections in projective
spaces in almost an identical fashion, for which the language of
GLSMs is absolutely appropriate.
} describes a complete
intersection in ${\mathbbm C}^{n}$ by a theory on
${\mathbbm C}^{n+k}$ with a superpotential, in which one adds
a new field (here, corresponding to the $y_j$)
for each hypersurface in the complete intersection.  
Furthermore, the cotangent complex is realized implicitly in
Yukawa couplings in the GLSM.  The superpotential for a 
complete intersection in ${\mathbbm P}^{n-1}$, for example, is
of the form
\begin{equation}
W \: = \: \sum_j p_j f_j(x),
\end{equation}
which has Yukawa couplings such as
\begin{equation}
\sum_{i,j}  \frac{\partial f_j}{\partial x_i}
\psi_-^i \psi_+^j.
\end{equation}
Following standard tricks, we can identify $\psi^i$ with $d x^i$,
and then the Yukawa coupling is a mass term that, in describing the
tangent bundle, realizes equation~(\ref{eq:cot-kernel}).
In this fashion we see these elementary aspects of derived geometry
appearing explicitly in standard GLSM constructions.
We will describe physical analogues of other constructions in the
next section.

Next, we will describe two more concrete examples of derived spaces.
The first is the derived critical locus, which is
defined as follows.  (See e.g. \cite{Vezzosi:1109.5213} for additional information.)
Let $X$ be a variety, and $W$ a holomorphic function
on $X$, whose critical locus is $Z$.  Then, the cotangent complex is
given by
\begin{equation}
0 \: \longrightarrow \: TX|_Z \: \stackrel{\partial^2 W}{\longrightarrow} \:
\Omega^1_X|_Z \: \longrightarrow \: 0,
\end{equation}
where $\Omega^1_X|_Z$ is in degree $0$ and
$TX|_Z$ is in degree $-1$.

Suppose that $X$ and $Z$ are smooth, then the complex above is
quasi-isomorphic to the one-element complex giving the cotangent bundle of
$Z$.  Let us outline this explicitly, at least for the special
case that $Z$ consists of (fat) points.
In this case, we dualize the short exact sequence
\begin{equation}
0 \: \longrightarrow \: TZ \: \longrightarrow \: TX|_Z \: 
\stackrel{\partial^2 W}{\longrightarrow} \: \Omega^1_X |_Z
\: \longrightarrow \: 0,
\end{equation}
and rewrite it as a quasi-isomorphism between complexes:
\begin{equation}
\xymatrix{
0 \ar[r] & TX|_Z \ar[d] \ar[r] &
\Omega^1_X |_Z \ar[r] \ar[d] & 0 \\
0 \ar[r] & 0 \ar[r] & \Omega^1_Z \ar[r] & 0.
}
\end{equation}
Thus, we see that in this special case,
the cotangent complex is quasi-isomorphic to $\Omega^1_Z$.

The second example of a derived manifold we shall encounter is the
derived zero locus.
Given a variety $X$, a vector bundle $E \rightarrow X$,
and a regular section $s \in \Gamma(E)$, the zero scheme
$Z \subset X$ of $s$ is a local complete intersection whose
cotangent complex is given by
\begin{equation}
0 \: \longrightarrow \: E^*|_Z \: \stackrel{ds}{\longrightarrow} 
\: \Omega^1_X |_Z
\: \longrightarrow \: 0,
\end{equation}
where $\Omega^1_X|_Z$ is in degree $0$, and $E^*|_Z$ is in degree $-1$,
and of course the codimension of $Z$ in $X$ equals the rank of $E$.

Suppose that $X$ is smooth, and the zero locus $Z$ is also smooth,
of codimension equal to the rank of $E$.
Then in this case, the cotangent complex is quasi-isomorphic to the
one-element complex giving the cotangent bundle of $Z$.  To see this,
we dualize the short exact sequence
\begin{equation}
0 \: \longrightarrow \: TZ \: \longrightarrow \: TX|_Z \: 
\stackrel{ds}{\longrightarrow}
\: E|_Z \: \longrightarrow \: 0
\end{equation}
and rewrite as a quasi-isomorphism between complexes:
\begin{equation}
\xymatrix{
0 \ar[r] & E^*|_Z \ar[r] \ar[d] & \Omega^1_X|_Z \ar[r] \ar[d] & 0 \\
0 \ar[r] & 0 \ar[r] & \Omega^1_Z \ar[r] & 0.
}
\end{equation}
Thus, we see that in this special case, the cotangent complex
is quasi-isomorphic to $\Omega^1_Z$.
(If the codimension of $Z$ is different from the rank of $E$,
then there is cohomology in degree $-1$, as we shall see in an
example in the next section.)

\subsection{Two-dimensional Landau--Ginzburg models}

In this section we review how 
a sigma model with a superpotential can be
interpreted in terms of derived manifolds.

Consider a two-dimensional (2,2) supersymmetric
Landau--Ginzburg model,
a nonlinear sigma model on $X$ with superpotential $W: X \rightarrow
{\mathbbm C}$.  Let us consider the tangent bundle arising in the
IR limit.  The Landau--Ginzburg theory has the Yukawa coupling term
\begin{equation}
\psi_+^i \psi_-^j D_i \partial_j W,
\end{equation}
so applying standard methods of (0,2) theories, if we let
$Z \equiv \{ {\rm d}W = 0 \}$ denote the critical locus, then at least
semi-classically, since the Yukawa coupling gives a mass to
elements of $TX|_Z$ that are not annihilated by\footnote{
In more detail,
the Yukawa coupling is defined by
\begin{equation}
D_i \partial_j W \: = \: \partial_i \partial_j W \: + \:
\Gamma_{ij}^k \partial_k W,
\end{equation}
but when we restrict to the critical locus $Z$, the second term drops out,
yielding just the ordinary Hessian.
}
\begin{equation}
D \partial W |_Z \: = \: \partial^2 W |_Z,
\end{equation}
the tangent bundle of the IR limit should be
described as the kernel of the map
\begin{equation}
TX|_Z \: \stackrel{\partial^2 W}{\longrightarrow} \:
\Omega^1_X|_Z.
\end{equation}

Mathematically, the complex above is the tangent complex of the
derived critical locus of $W$.  The cotangent complex is the dual
complex, namely
\begin{equation}  \label{eq:l-der-crit}
TX|_Z \: \stackrel{\partial^2 W}{\longrightarrow} \:
\Omega^1_X|_Z.
\end{equation}

Let us consider a special case.  Suppose $X$ is the total space of
a vector bundle $V \rightarrow M$, and $W = p f$ where
$f \in \Gamma(V)$ and $p$'s are fiber coordinates on $V^*$,
so that in the IR this theory should flow to a nonlinear sigma model
on $Z' = \{f=0\} \subset M$ (for essentially the same reasons as in
analysis of large-radius limits of GLSMs).
Let us also assume that $Z'$ is smooth,
so that $Z'$ coincides with the critical locus $Z$.
In this case, the restriction of the Hessian $D \partial W$ to the critical
locus is of the form
\begin{equation}
\left[ \begin{array}{cc}
0 & \partial_i f \\
\partial_j f & 0
\end{array} \right].
\end{equation}
Thus, in this case, the mass matrix defined by the Yukawa coupling
is describing the cotangent complex~(\ref{eq:l-der-crit}).

There is also an analogue of these considerations for (0,2) supersymmetric
theories.  Consider a (0,2) Landau--Ginzburg model, defined by a space
$X$, holomorphic vector bundle ${\cal E} \rightarrow X$ satisfying the
anomaly cancellation condition 
\begin{equation}
{\rm ch}_2({\cal E}) \: = \: {\rm ch}_2(TX),
\end{equation}
and $J \in \Gamma({\cal E}^*)$ defining
a (0,2) superpotential\footnote{
In general, a (0,2) theory has potentials defined by both $J \in \Gamma(
{\cal E}^*)$ as well as $E \in \Gamma( {\cal E})$, satisfying
$E \cdot J = 0$, but in this case for simplicity we restrict to the
special case $E \equiv 0$.
}.  
The Yukawa couplings
\begin{equation}
\psi_+^i \lambda_-^a D_i J_a
\end{equation}
(where $\psi_+^i$s are right-moving fermions and $\lambda_-^a$s are left-moving
fermions)
define a mass matrix that implies the
left-movers couple to the kernel of
\begin{equation}
{\cal E}|_Z \: \stackrel{\nabla J}{\longrightarrow} \: \Omega^1_{Z/X}
\end{equation}
where $Z \equiv \{J=0\}$,
and the right-movers couple to the cokernel.
(Fermions not in either the kernel or cokernel get a mass, and so are
integrated out along RG flow.)

It will be helpful to consider a concrete example arising in
two-dimensional (0,2) theories.  Suppose we wish to describe an
IR theory on a 
space $Z$ with bundle ${\cal E}$ given as the kernel
\begin{equation}
0 \: \longrightarrow \: {\cal E} \: \longrightarrow \:
M \: \stackrel{F}{\longrightarrow} \: L \: \longrightarrow \: 0,
\end{equation}
where $M$, $L$ are holomorphic vector bundles on $B$.
This theory arises as the IR endpoint of a (0,2) Landau--Ginzburg
model on a space
\begin{equation}
X \: \equiv \: {\rm Tot}\left( \pi: \, L^* \: \longrightarrow \: B \right),
\end{equation}
with (0,2) superpotential
\begin{equation}
W \: = \: p \Lambda^a \pi^* F_a,
\end{equation}
where the $\lambda^a$ are Fermi superfields coupling to the bundle $M$,
and $p$ is a fiber coordinate on $L^*$.
The $F_a$ are simply indexed components of the map $F: M \rightarrow L$.
The (0,2) superpotential is defined by
$J_a = p \pi^* F_a$.
In principle, the theory should flow in the IR to the zero locus of
$J$ (just as in a (2,2) theory, a Landau--Ginzburg model flows to the
critical locus of the superpotential).
In this case, since $F$ is surjective, the zero locus is $B = \{ p = 0 \}$.
Note that $J \in \Gamma(M^*)$, and so we define $E = M^*$.
The cotangent complex of the derived zero locus of $J$ is given by
\begin{equation}
E^* |_B \: \stackrel{{\rm d}J}{\longrightarrow} \: \Omega^1_X |_B
\end{equation}
and since the codimension of $B$ is different from the rank of $E$ in
general, there is nonzero cohomology in both degree $0$ and $-1$.
Specifically, the cohomology at degree $0$ is
$\Omega^1_B$, and the cohomology at degree $-1$ is 
${\cal E}$.
Thus, we see that the cotangent complex of the derived zero locus
is encoded in the physics of two-dimensional (0,2) Landau--Ginzburg
models.

\subsection{Derived structures on moduli spaces of SCFTs}

So far we have discussed two-dimensional Landau--Ginz\-burg models as
giving a physical realization of derived schemes.  Next, we turn our
attention to moduli spaces of SCFTs, much as we did in 
Section~\ref{sect:mod-ell-curves}
in our examination of stacks, as would arise in four-dimensional 
$N=1$ supergravity theories obtained from string compactification.
We will outline how derived structures on such moduli spaces
(specifically, the tangent and cotangent complexes)  seem
to be encoded in worldsheet physics, and outline how cohomology at nonzero
degree corresponds to singular points and enhanced gauge symmetries. 
(In other words, if the cohomology of the cotangent complex
at degree $0$ corresponds to scalars in the target-space theory --
infinitesimal moduli of the compactification -- then cohomology
at degree $-1$ corresponds to vectors arising at enhanced symmetry
points.)

In four dimensions, it is
important to note that a derived structure on
the moduli space cannot be described as a holomorphic derived critical
locus of the spacetime superpotential, simply because the superpotential
is a section of a line bundle \cite{Witten:1982hu}, 
the line bundle of holomorphic top-forms
(often called the Hodge line bundle, and identified with the tensor square
of the Bagger--Witten line bundle).
As a result, the critical locus is defined by not only the superpotential,
but also a choice of connection on the Bagger--Witten line bundle 
(defined physically
by the K\"ahler potential).  Although the superpotential itself is
holomorphic, the connection is not, and so holomorphic derived geometry
cannot be relevant here.

To see, for example, the cotangent complex, it is difficult to work
directly with states in a nonlinear sigma model on a Calabi--Yau,
simply because in the case that the complex has cohomology at degree $-1$,
the Calabi--Yau is singular, and so the nonlinear sigma model becomes
ill-behaved.  It may still be possible to work directly with such
singular theories, but we will take a different approach.  Instead of 
working with IR nonlinear sigma models, we will work with UV theories,
which will sidestep this issue.

In general terms, we will identify BRST complexes of states
with the cotangent complex. From a more global perspective,
we are proposing a shift in emphasis from describing states and operators
in terms of BRST cohomology, to BRST complexes of states,
without taking cohomology.
(See for example \cite{Thomas:1999ic} for a similar appeal in a different
context.)  In general terms, we expect that RG flow preserves BRST
cohomology, so we expect that BRST complexes of states in theories related
by RG flow, are quasi-isomorphic to one another.  A specific choice of
renormalization scheme should define a particular quasi-isomorphism.
In any event, we will look for cotangent complexes on moduli spaces
of SCFTs by looking at structures in UV theories.

To be concrete, we will outline this notion in an example,
a heterotic
compactification on a Landau--Ginzburg orbifold corresponding to the
(2,2) quintic, with superpotential
\begin{equation}
W \: = \: \frac{1}{5} \sum_{i_1 \cdots i_5} w_{i_1 \cdots i_5}
\Phi^{i_1} \cdots \Phi^{i_5},
\end{equation}
as described in \cite{Kachru:1993pg}.
To be concrete, let us consider the ($E_6$ singlet) moduli appearing in the
$k=1$ sector of the orbifold, as described in \cite{Kachru:1993pg}[Section 3.2].
As discussed there, the moduli -- the four-dimensional scalars -- arise
on the worldsheet as the cohomology
at $q_+ = -1/2$ in the complex
\begin{equation}
0 \: \longrightarrow \: V_{-3/2} \: \stackrel{ \overline{Q}_{+,L} }{
\longrightarrow} \: V_{-1/2} \: \longrightarrow \: 0
\end{equation}
where 
\begin{eqnarray}
\overline{Q}_{+,L} & = &
\psi^{i_1}_{2/5} w_{i_1\cdots i_5}
\phi^{i_2}_{-1/10} \cdots \phi^{i_5}_{-1/10}\,+
\nonumber \\
& &  \: + \:
4 \psi^{i_1}_{-3/5} w_{i_1\cdots i_5} \phi^{i_2}_{9/10}\phi^{i_3}_{-1/10}
\cdots \phi^{i_5}_{-1/10}
\end{eqnarray}
forms the realization\footnote{
Strictly speaking, the authors of \cite{Kachru:1993pg} are using a spectral sequence
to compute BRST cohomology, so $\overline{Q}_{+,L}$ is only part of the
BRST operator, but the rest has already been taken into account, so without
loss of generality we may as well consider this to be the BRST operator.
} of the BRST operator, the states $V_{-1/2}$
at $q_+ = -1/2$ are of the form
\begin{equation}
\phi^{i_1}_{-1/10} \cdots \phi^{i_4}_{-1/10} \psi^j_{-3/5} | 0 \rangle,
\end{equation}
and the states $V_{-3/2}$ at $q_+=-3/2$ are of the form
\begin{equation}
\overline{\psi}_{-2/5, i} \psi^j_{-3/5} | 0 \rangle, \: \: \:
\overline{\phi}_{-9/10, j} \phi^i_{-1/10} | 0 \rangle.
\end{equation}

The cohomology at $V_{-1/2}$ corresponds to gauge singlet chiral multiplets
in the four-dimensional compactification -- spacetime scalars, 
infinitesimal moduli of the SCFT, 
in other
words.  The cohomology at $V_{-3/2}$ corresponds
to extra $U(1)$'s, extra vector multiplets, in the four-dimensional
compactification.
Working through the details, one finds that for the Fermat quintic,
the cohomology at degree $-1/2$ has dimension 305.  As discussed in
\cite{Kachru:1993pg}, the states at degree $-1/2$
are represented by five quartic functions
$P_i(\phi_{-1/10})$ subject to the relation
\begin{equation}
P_i \: \sim \: P_i \: + \: A_i^j \frac{\partial W}{\partial \phi_j}
\: + \: \phi^k B_{k \ell} \frac{\partial^2 W}{\partial \phi^{\ell} 
\partial \phi^i},
\end{equation}
where $W$ is the superpotential, and $A$, $B$ are arbitrary constant
matrices.

The cohomology at
degree $-3/2$ has dimension 5.  Of those five elements of cohomology at
degree $-3/2$, one state is present for generic complex structures.

We can understand these states somewhat more systematically as follows.
Since we are on the (2,2) locus, 
albeit at a Landau--Ginzburg orbifold
point, 
morally we expect these singlets to be related to complex
structure moduli $H^1(T)$, K\"ahler moduli $H^1(T^*)$, and bundle
moduli $H^1({\rm End}\, T)$.  (Rather, these are the moduli\footnote{
Off the (2,2) locus, only a subset of the complex and bundle moduli
will be present in the CFT in general, because not all complex structure
deformations are compatible with a given holomorphic bundle.
See e.g. \cite{Anderson:2011ty,Melnikov:2011ez} for recent discussions.
} in the
corresponding large-radius (2,2) supersymmetric
nonlinear sigma model, so barring massing up of
pairs, one expects a similar, though not necessarily identical, counting
of moduli at Landau--Ginzburg.)

In this language, following \cite{Kachru:1993pg},
we can understand the complex structure deformations
as states of the form
\begin{equation}
P_i(\phi_{-1/10}) \psi^j_{-3/5} | 0 \rangle,
\end{equation}
where $P_i = \partial_i S$ for some quintic polynomial $S$, subject
to the equivalence
\begin{equation}
S \: \sim \: S \: + \: \phi^i A_i^j \partial_j W \: + \:
\phi^i \phi^k B_k^{\ell} \frac{ \partial^2 W }{\partial \phi^{\ell}
\partial \phi^i}.
\end{equation}

Similarly, the analogue of bundle deformations is encoded \cite{Kachru:1993pg}
in the space
of five quartic polynomials $P_i$ such that $\phi^i P_i = 0$,
subject to the equivalence relation
\begin{equation}
P_i \: \sim \: P_i \: + \: A_i^j \partial_j W \: - \: \frac{1}{5}
\partial_i\left( \phi^k A_k^j \partial_j W \right).
\end{equation}
For special values of the superpotential $W$, this equivalence relation
is less powerful, so that the Landau--Ginzburg theory has extra
$E_6$ singlets.  For example, the Fermat quintic has five extra states
\begin{equation}
\left( \frac{1}{4} \phi^i_{-1/10} \overline{\phi}_{-9/10, i} \: - \:
\overline{\psi}_{-2/5, i} \psi^i_{-3/5} \right) | 0 \rangle.
\end{equation}
In this case, there are extra gauge bosons arising at $q_+ = -3/2$
and extra scalars at $q_+ = -1/2$.
In the target-space supergravity theory, as one moves away from
these special points, the extra gauge bosons are Higgsed.

For our purposes, the essential point is that these BRST complexes
of states have the structure of a cotangent complex,
indicating a physically-meaningful derived structure on the moduli space of
SCFTs.  In this realization,
cohomology of the cotangent complex at degree $-1$ (here, charge $q_+ = -3/2$)
corresponding to extra gauge bosons in the target-space theory,
in accord with standard lore that at special/singular points in the
moduli space, the target-space theory has enhanced gauge symmetries.

\subsection{Derived stacks}

We have outlined in this section how at least certain derived schemes
appear to have a physical realization in terms of Landau--Ginzburg models,
and we have also seen that certain (Deligne--Mumford) stacks have
a physical realization in terms of gauged sigma models.  With this in mind,
the physical realization of certain derived stacks should be as gauged
Landau--Ginzburg models or gauged sigma models with superpotential, 
of which the most common examples are
gauged linear sigma models \cite{Witten:1993yc}.

\section{Conclusions}

In this paper we have outlined three examples of physical realizations
of mathematical structures in which renormalization group flow
realizes categorical or homotopy equivalences:  stacks, derived categories,
and derived schemes.

The fact that renormalization group flows seems to often realize
categorical or homotopy equivalences would appear to suggest that there
may be some way to define a model structure on a category of
renormalization group flows, such that weak equivalences relate theories
which RG flow to the same endpoint.
We have not so far succeeded in finding completely sensible
definitions, but in general terms, one can outline an idea.  Let
${\cal C}$ be a category of renormalization-group flows of `one'
theory, meaning that the objects are quantum field theories at
various scales and the morphisms correspond to RG flows toward
lower energies.  One expects that such
a category should have an initial object (a UV
fixed point, which should be a cofibrant object in the sense of
model categories) 
and a terminal object (an IR fixed point, which should be a fibrant
object in the sense of model categories).  One also expects
it will have pushouts and pullbacks.  For example, if a field theory
$A$ RG flows to both theories $B$ and $C$ at different scales,
then they should flow to a common theory at a lower scale, so in a nutshell
one expects pushouts to exist, and for similar reasons, one also expects
pullbacks to exist.  Now, if we want to associate weak equivalences
with maps between theories related by RG flow, we have a minor issue:
the morphisms only run from higher to lower energies.  We can solve this
by localizing on the morphisms.  In other words, if $S$ denotes the
set of morphisms, we could consider the category
$S^{-1} {\cal C}$, and then associate weak equivalences with all
morphisms.  It is less clear how one should define fibrations and
cofibrations in this category, however, in a way that is both nontrivial
and satisfies axiom MC5 of \cite{dwyer1995homotopy}.

\bibliography{allbibtex}

\bibliographystyle{prop2015}

\end{document}